\def\nk{n_{\rm b}}

\def\rfr#1{Equation\,(\ref{#1})}
\def\rfrs#1#2{Equations\,(\ref{#1})--(\ref{#2})}
\def\Rfr#1{Equation\,(\ref{#1})}

\def\dert#1#2{\frac{{{\textrm{d}}}{#1}}{{{\textrm{d}}}{#2}}}

\def\virg#1{``#1"}

\def\eqi{\begin{equation}}
\def\eqf{\end{equation}}
\def\eqia{\begin{eqnarray}}
\def\eqfa{\end{eqnarray}}

\def\rp#1#2{{#1\over#2}}
\def\lb#1{\label{#1}}

\def\bds#1{\boldsymbol{#1}}

\def\ton#1{\left(#1\right)}
\def\qua#1{\left[#1\right]}
\def\grf#1{\left\{#1\right\}}

\documentclass[onecolumn]{aastex}

\usepackage{morefloats}
\usepackage[title]{appendix}
\usepackage{hyperref,textcomp}
\usepackage{booktabs}
\usepackage[table,xcdraw]{xcolor}
\usepackage{multirow}
\usepackage{rotating,tabularx}
\usepackage{float}
\usepackage{enumerate}
\usepackage{rotating}
\usepackage[polutonikogreek,english]{babel}
\usepackage{amsmath,starfont,textgreek,w-greek,wasysym}
\usepackage[flushleft]{threeparttable}
\usepackage{amsthm}
\usepackage{amscd,lineno}
\usepackage{amssymb,dsfont}
\usepackage{graphicx,epsfig}
\usepackage{txfonts}
\usepackage{xr-hyper}

\RequirePackage{color}

\newcommand{\emaila}{lorenzo.iorio@libero.it}

\usepackage{hyperref}

\hypersetup{
    colorlinks=true,
    linkcolor=red,
    citecolor=magenta,
    filecolor=magenta,
    urlcolor=cyan,
    pdftitle={Modified Potentials},
    pdfpagemode=FullScreen,}

\allowdisplaybreaks[1]

\begin{document}

\title{Effect of some modified models of gravity on the radial velocity of binary systems}

\shortauthors{L. Iorio, M.L. Ruggiero}

\author{Lorenzo Iorio\altaffilmark{1} }
\affil{Ministero dell'Istruzione, dell'Universit\`{a} e della Ricerca
(M.I.U.R.)
\\ Viale Unit\`{a} di Italia 68, I-70125, Bari (BA),
Italy}

\email{\emaila}

\author{Matteo Luca Ruggiero}
\affil{Dipartimento di Matematica ``G.Peano'', Universit\`a degli studi di Torino, Via Carlo Alberto 10, 10123 Torino, Italy. \\ INFN - LNL , Viale dell'Universit\`a 2, 35020 Legnaro (PD), Italy}
\email{matteoluca.ruggiero@unito.it}

\begin{abstract}
For many classes of astronomical and astrophysical binary systems, long observational records of their radial velocity $V$, which is their directly observable quantity, are available. For exoplanets close to their parent stars, they cover several full orbital revolutions, while for wide binaries like, e.g., the Proxima/$\alpha$ Centauri AB system, only relatively short orbital arcs are sampled by existing radial velocity measurements. Here, the changes $\Delta V$ induced on a binary's radial velocity by some long-range modified models of gravity are analytically calculated. In particular,  extra-potentials proportional to $r^{-N},\,N=2,\,3$ and $r^2$ are considered; the Cosmological Constant $\Lambda$ belongs to the latter group. Both  the net shift per orbit and the instantaneous one are explicitly calculated for each model. The Cosmological Constant induces a shift in the radial velocity of the Proxima/$\alpha$ Centauri AB binary as little as $\left|\Delta V\right|\lesssim 10^{-7}\,\mathrm{m\,s}^{-1}$, while the present-day accuracy in measuring its radial velocity is $\sigma_V\simeq 30\,\mathrm{m\,s}^{-1}$. The calculational scheme presented here is quite general, and can be straightforwardly extended to any other modified gravity.
\end{abstract}


\keywords{Gravitation -- Celestial mechanics -- stars: binaries: spectroscopic -- stars: planetary systems}
\section{Introduction}
Although general relativity, after more than one century since its birth, has always passed all the experimental and observational tests devised so far to put it to the test in various scenarios \citep{2014LRR....17....4W,2016Univ....2...22C,2019Univ....6....9C,2020Univ....6..154D,2020Univ....6..156W}, there is an increasingly rich phenomenology, at both astrophysical (dark matter \citep{2017ARA&A..55....1F,2018ARA&A..56..435W,2019IJMPA..3430013K}) and cosmological (dark energy \citep{2008ARA&A..46..385F,2018RPPh...81a6902B}) scales, pointing towards the\textcolor{black}{\footnote{\textcolor{black}{It should be stressed that most of the current research in the field of dark matter and dark energy is, actually, made within general relativity. Of course, there is the option of alternative theories of gravity, but it would be incorrect to look at them as a necessity from the point of view of dark matter and dark energy.}}} \textcolor{black}{option} of, perhaps, modifying it in such long-range domains; for a recent overview, see, e.g. \citep{2016Univ....2...23D}, and references therein. To cope with such potential difficulties of the Einstein's theory, several alternative models of the gravitational interaction have been devised so far; see, e.g., \citet{2007IJGMM..04..115N,Lobo09,2011PhRvD..84b4020H,2012PhR...513....1C,2015Univ....1..199C,Berti:2015itd,2016RPPh...79j6901C,2017PhR...692....1N,universe7080269}, and references therein. A major drawback of all such theoretical schemes is that, to date, no independent tests exist for them other than just the phenomena for which they were introduced at the time. Thus, devising alternative ways to empirically scrutinize it, at least in principle, in different arenas is quite important.

Several long-range modified models of gravity envisage power-law modifications of the $r^{-1}$ Newtonian potential of a central body  proportional to $r^{-N},\,N>1$, where $r$ is the distance from it. They induce deviations from the inverse-square law in terms of small additional accelerations proportional to $r^{-N-1},\, N>1$. Such kind of effects are better constrained with tight binary systems like, e.g., several exoplanets \citep{2010exop.book.....S,2018haex.bookE....D,2018exha.book.....P} many of which orbit at $r\simeq 10^{-3}$ astronomical units (au) from their parent stars. One of the most widely adopted observable in detecting them is the radial velocity\footnote{To date, according to the online database http://exoplanet.eu/, about a thousand planets have been discovered with the RV method.} (RV) $V$ of the reflex motion of their host stars displaced by the gravitational tug of the planets \citep{2016PASP..128f6001F,2018haex.bookE...4W,2020ApJ...905..155G,2022AJ....163...63M}, i.e. the projection of the barycentric stellar velocity vector ${\mathbf{v}}_\star$ onto the line of sight, even though other observables can be in principle be used (see e.g. \citet{Ruggiero:2020yoq} and references therein). It is expected that the accuracy in measuring exoplanets' RV  may be pushed, at least in principle, to $0.2-0.5\,\mathrm{m\,s}^{-1}$ \citep{2020ApJ...905..155G}, or even down to the $0.01\,\mathrm{m\,s}^{-1}$ level \citep{2022AJ....163...63M}. To the best knowledge of the present authors, no calculations of the impact of the aforementioned modified models of gravity on the RV have been performed in the literature so far. This paper aims to fill this gap by analytically calculating the shifts $\Delta V$ of the RV due to some of the most widely discussed $r^{-N}$ extra-potentials, i.e., those with $N=2,\,3$. Both the instantaneous and the orbit averaged RV variations are computed; the latter ones are particularly suitable for this type of planets since for most of them long data records covering many orbital revolutions exist.

Also extremely wide binaries for which RV data exist \citep{1990AJ....100.1968C,2017A&A...598L...7K}, like, e.g., Proxima orbiting the pair $\alpha$ Centauri AB in more than $5\times 10^5\,\mathrm{yr}$ \citep{2017A&A...598L...7K}, may turn out to be useful, at least in principle, to put to the test another class of modified models of gravity whose extra-potentials go as $r^2$. In particular, the Cosmological Constant\footnote{The CC is the most straightforward explanation within general relativity  for the paradigm phenomenologically dubbed as \virg{dark energy} driving the observed late-time cosmic acceleration \citep{1998AJ....116.1009R,1999ApJ...517..565P}.} (CC) $\Lambda$ \citep{1989RvMP...61....1W,1992ARA&A..30..499C,2001LRR.....4....1C,2003RvMP...75..559P,2003PhR...380..235P,2004sgig.book.....C,Davis:2010,2018EPJH...43...73O} induces a small extra-acceleration which is proportional to $r$ \citep{2001rsgc.book.....R,Kerr03}.
On the other hand,  also several classes of long range modified models of gravity aiming to explain in a unified way seemingly distinct features of the cosmic dynamics like early-time inflation, late-time acceleration driven by dark energy and even dark matter imply a CC-type parameterization \citep{Allemandi:2005tg,2007IJGMM..04..115N,Allemandi:2006bm,2007JPhA...40.6725N,2010PhRvD..82b3519D,2010LRR....13....3D,
2011PThPS.190..155N,2011PhR...509..167C,2012PhR...513....1C,2012AnP...524..545C,2015Univ....1..199C,2015Univ....1..123D,
2015IJMPD..2441002C,2016RPPh...79j6901C,Iorio:2016sqy}.
Since it is of the utmost importance to try to independently test the CC in different scenarios with respect to the cosmological ones which, only they, have justified its introduction to date, attempting to use such wide binaries and their RV measurements to tentatively constrain $\Lambda$ should be deemed as a valuable effort. Thus, also the RV change $\Delta V$ due to the Hooke-like acceleration induced by $\Lambda$ is explicitly worked out. Also in this case, both the instantaneous and the orbit averaged shifts are analytically calculated. Nonetheless, only the former one can be of practical use since the currently available RV records of wide binaries do not cover a full orbital period for none of them.
\section{The calculational scheme}\lb{sec:2}
In order to set up the calculational scheme for computing the shift $\Delta V$ of the RV $V$ induced by any small perturbing acceleration $\bds A$ with respect to the Newtonian monopole, we will follow \citet{1993CeMDA..55..209C}. All the following results hold for the binary's relative orbit; the resulting shift $\Delta V$ for the stellar RV can be straightforwardly obtained by rescaling the final formula by the ratio of the planet's mass $M_\mathrm{p}$ to the sum $M\doteq M_\star + M_\mathrm{p}$ of the masses of the parent star and of the planet itself.

The velocity vector $\mathbf{v}$ is
\eqi
\mathbf{v} = \mathrm{v}_R\,{\mathbf{u}}_R +  \mathrm{v}_T\,{\mathbf{u}}_T +  \mathrm{v}_N\,{\mathbf{u}}_N,
\eqf
where
\begin{align}
{\mathbf{u}}_R \lb{uR}& = \ton{\cos\Omega\,\cos u - \cos I\,\sin\Omega\,\sin u}\,\mathbf{i}+\ton{\sin\Omega\,\cos u + \cos I\,\cos\Omega\,\sin u}\,\mathbf{j}+ \sin I\,\sin u\,\mathbf{k}, \\ \nonumber \\
{\mathbf{u}}_T \lb{uT}& =\ton{-\sin u\,\cos\Omega - \cos I\,\sin\Omega\,\cos u}\,\mathbf{i} + \ton{-\sin\Omega\,\sin u + \cos I\,\cos\Omega\,\cos u}\,\mathbf{j} +  \sin I\,\cos u\,\mathbf{k}, \\ \nonumber \\
{\mathbf{u}}_N \lb{uN}& = \sin I\,\sin\Omega\,\mathbf{i} -\sin I\,\cos\Omega\,\mathbf{j} + \cos I\,\mathbf{k},
\end{align}
are the unit vectors along the radial, transverse and normal directions, respectively, of the trihedron co-moving with the test particle \citep{1991ercm.book.....B}. In \rfrs{uR}{uN}, $I$ is the inclination of the orbit to the reference $\grf{x,\,y}$ plane, $\Omega$ is the longitude of the ascending node, and $u\doteq \omega + f$ is the argument of latitude given by  the sum of the argument of pericentre $\omega$ and the true anomaly $f$.

If a small perturbing acceleration $\bds A$ is present, the velocity $\mathbf{v}$ is, in general, changed by an amount \citep{1993CeMDA..55..209C}
\eqi
\Delta\mathbf{v} =\Delta\mathrm{v}_R\,{\mathbf{u}}_R +  \Delta\mathrm{v}_T\,{\mathbf{u}}_T +  \Delta\mathrm{v}_N\,{\mathbf{u}}_N ,
\eqf
where \citep{1993CeMDA..55..209C}
\begin{align}
\Delta\mathrm{v}_R \lb{DvR}& = -\rp{\nk\,a\,\sin f}{\sqrt{1-e^2}}\,\ton{\rp{e}{2\,a}\,\Delta a + \rp{a}{r}\,\Delta e} - \rp{\nk\,a^3}{r^2}\,\Delta\mathcal{M} -\rp{\nk\,a^2}{r}\,\sqrt{1-e^2}\ton{\cos I\,\Delta\Omega + \Delta\omega}, \\ \nonumber \\
\Delta\mathrm{v}_T \lb{DvT}& = -\rp{\nk\,a\,\sqrt{1-e^2}}{2\,r}\,\Delta a + \rp{\nk\,a\,\ton{e + \cos f}}{\ton{1-e^2}^{3/2}}\,\Delta e +  \rp{\nk\,a\,e\,\sin f}{\sqrt{1-e^2}}\,\ton{\cos I\,\Delta\Omega + \Delta\omega}, \\ \nonumber \\
\Delta\mathrm{v}_N \lb{DvN}& = \rp{\nk\,a}{\sqrt{1-e^2}}\,\qua{\ton{\cos u + e\,\cos\omega}\,\Delta I + \sin I\,\ton{\sin u + e\,\sin\omega}\,\Delta\Omega}.
\end{align}
In \rfrs{DvR}{DvN}, $a$ is the semimajor axis, $\nk\doteq\sqrt{\mu/a^3}$ is the Keplerian mean motion, $\mu\doteq GM$ is the binary's gravitational parameter, $G$ is the Newtonian constant of gravitation, $e$ is the eccentricity, and $\mathcal{M}$ is the mean anomaly. The shifts $\Delta a,\,\Delta e,\,\Delta I,\,\Delta\Omega,\,\Delta\omega,\,\Delta\mathcal{M}$ are to be meant as instantaneous, i.e. they are functions of time through some of the time-dependent anomalies connected with the position of the test particle along its orbit.
The variation $\Delta\mathcal{M}$ of the mean anomaly $\mathcal{M}$ must be calculated as \citep{2019EPJC...79..816I}
\eqi
\Delta\mathcal{M}  = \Delta\eta+\int_{t_0}^t\Delta\nk\ton{t^{'}}\mathrm{d}t^{'},\lb{DMgeneral}
\eqf
where $\eta$ is the mean anomaly at epoch, and \citep{2019EPJC...79..816I}
\eqi
\int_{t_0}^t\Delta\nk\ton{t^{'}}\mathrm{d}t^{'} = -\int_{q_0}^q\rp{3}{2}\,\rp{\nk}{a}\,\Delta a\ton{q^{'}}\,\dert{t}{q^{'}}\mathrm{d}q^{'};\lb{konz}
\eqf
in \rfr{konz}, $q$ denotes the time-dependent anomaly, like the true anomaly $f$ or the eccentric anomaly $E$, specifically chosen as fast variable of integration.

The shift $\Delta V$ of the RV is calculated from the component $\Delta{\mathbf{v}}_z$ along the reference $z$ axis which is customarily aligned with the line-of-sight, while the $\grf{x,\,y}$ plane coincides with the plane of the sky.
\section{The case of  a Hooke-type acceleration}\lb{sec:3}
Here, we treat the RV shift induced by a perturbing radial acceleration proportional to the distance $r$ \citep{2001rsgc.book.....R,Kerr03,2007PhRvD..75h2001A} given by
\eqi
{\bds A}_{\mathcal{K}} = \mathcal{K}\,r\,{\mathbf{u}}_R\textcolor{black}{.} \lb{Acc}
\eqf
In the case of \rfr{Acc}, it is computationally more convenient to adopt the eccentric anomaly $E$  in terms of which the following Keplerian relations are expressed
\begin{align}
r & = a\ton{1-e\,\cos E}, \\ \nonumber \\
\sin f & = \rp{\sqrt{1-e^2}\,\sin E}{1-e\,\cos E}, \\ \nonumber \\
\cos f & = \rp{\cos E -e}{1-e\,\cos E}, \\ \nonumber \\
\dert{t}{E} \lb{dtdE}& = \rp{1-e\,\cos E}{\nk}.
\end{align}
The instantaneous shifts of any  orbital element $\kappa$
has to be calculated as
\eqi
\Delta\kappa\ton{E} = \int_{E_0}^E\dert{\kappa}{t}\dert{t}{E^{'}}\mathrm{d}E^{'},\,\kappa=a,\,e,\,I,\,\Omega,\,\omega,\,\eta,
\eqf
where $\mathrm{d}\kappa/\mathrm{d}t$ are given by the Gauss equations for the rates of change of the orbital elements, and $\mathrm{d}t/\mathrm{d}E$ is given by \rfr{dtdE}.

By using \rfr{Acc}, one has
\begin{align}
\Delta a\ton{E} \lb{Da} & = -\rp{\mathcal{K}\,a^4 e\,\ton{\cos E_0-\cos E}\,\qua{-2+e\,\ton{\cos E_0 + \cos E}}}{\mu}, \\ \nonumber \\
\Delta e\ton{E} \lb{De} & = \rp{\mathcal{K}\,a^3\,\ton{-1 + e^2}\,\ton{\cos E_0-\cos E}\,\qua{-2+e\,\ton{\cos E_0 + \cos E}}}{2\,\mu}, \\ \nonumber \\
\Delta I\ton{E} \lb{DI}& = 0, \\ \nonumber \\
\Delta\Omega\ton{E} \lb{DO}& = 0, \\ \nonumber \\
\Delta\omega\ton{E} \lb{Domega}& = \rp{\mathcal{K}\,a^3\,\sqrt{1-e^2}\,\grf{4\,\ton{1+e^2}\,\ton{\sin E_0-\sin E} - e\,\qua{6\,\ton{E_0-E} +\ton{\sin 2E_0 - \sin 2E}} }}{4\,e\,\mu}, \\ \nonumber \\
\Delta\eta\ton{E} \lb{Deta}& = \nonumber \rp{\mathcal{K}\,a^3}{12\,e\,\mu}\,\qua{ 6\,e\,\ton{7 + 3\,e^2}\,\ton{E_0 - E} - 6\,\ton{2 + 12\,e^2 +\,e^4}\,\ton{\sin E_0 -\sin E}  + \right. \\ \nonumber \\
&\left. +  3\,e\,\ton{1 + 5\,e^2}\,\ton{\sin 2 E_0-\sin 2 E}  - 2\,e^4\,\ton{\sin 3 E_0 - \sin 3E}}.
\end{align}

From \rfr{dtdE} and \rfr{Da}, it turns out that \rfr{konz} yields
\begin{align}
\int_{t_0}^t\Delta\nk\ton{t^{'}}\mathrm{d}t^{'} & \lb{Intdn}\nonumber = \rp{\mathcal{K}\,a^3\,e}{8\,\mu}\,\grf{
12\,\qua{e\,\ton{E_0 - E} - 2\,\ton{\sin E_0  - \sin E }} + 24\,\cos E_0\,\ton{E_0 - E + e\,\sin E} + \right.\\ \nonumber \\
\nonumber &\left. +  e\,\qua{-6\,\cos 2 E_0  \ton{E_0 - E +\,e\,\sin E} - 3\,\ton{\sin 2 E_0  + 3\,\sin 2 E} + \right.\right.\\ \nonumber \\
&\left.\left. + e\,\ton{-8\,\sin^3 E_0  + 3\,\sin E  +\,\sin 3 E}}
},
\end{align}
so that, from \rfr{DM} calculated with \rfr{Deta} and \rfr{Intdn}, one finally has
\begin{align}
\Delta\mathcal{M}\ton{E} \nonumber \lb{DM}& = \rp{\mathcal{K}\,a^3}{24\,e\,\mu}\,\grf{
12\,e\,\ton{7 + 6\,e^2}\,\ton{E_0 - E} - 4\,\ton{6 + 54\,e^2 + 7\,e^4}\,\sin E_0  + \right.\\ \nonumber \\
\nonumber &\left. +  6\,e\,\sin 2 E_0  + 3\,\ton{8 + 72\,e^2 + 7\,e^4}\,\sin E  +\right.\\ \nonumber \\
 \nonumber &\left. + 2\,e^3\,\cos 2 E_0\,\ton{-9 E_0 + 9 E + 2 e\,\sin E_0  - 9 e\,\sin E } + \right.\\ \nonumber \\
\nonumber &\left. +  6\,e^2\,\cos E_0\,\qua{7 e\,\sin E_0  + 12\,\ton{E_0 - E + e\,\sin E }} -\right.\\ \nonumber \\
&\left. -  3\,e\,\ton{2 + 19\,e^2}\,\sin 2 E  + 7\,e^4\,\sin 3 E
}.
\end{align}

By inserting \rfrs{Da}{Domega} and \rfr{DM} into \rfrs{DvR}{DvN} allows one to obtain an exact expression for the instantaneous shift $\Delta V\ton{E}$ of the radial velocity induced by \rfr{Acc}; it is too cumbersome to be displayed explicitly here. Below, we show an expansion of it to the first order in the eccentricity $e$, which reads
\begin{align}
\Delta V\ton{E} \lb{DV1}& = \rp{\mathcal{K}\,a^{5/2}\,\sin I}{\sqrt{\mu}}\,\mathcal{V}\ton{E},
\end{align}
with
\begin{align}
\mathcal{V}\ton{E}\nonumber & = \cos\ton{E_0 - 2E - \omega} - \cos\ton{E + \omega} +
 2\,\ton{-E_0 + E}\,\sin\ton{E + \omega} + \\ \nonumber \\
\nonumber &+ \rp{e}{4}\grf{
5\,\cos\ton{E_0 - 3 E - \omega} - \cos\ton{2 E_0 - 2 E - \omega} +
 17\,\cos\ton{E_0 - E - \omega} - \right.\\ \nonumber \\
\nonumber &\left. -  15\,\cos\omega - \cos\ton{E_0 - E + \omega} - 21\,\cos\ton{E_0 + E + \omega} +
 16\,\cos\ton{2 E + \omega} - \right.\\ \nonumber \\
\nonumber &\left. -  2\,\ton{E_0 - E}\,\qua{2\,\sin\omega + 6\,\cos E_0\,\sin\ton{E + \omega} +
    9\,\sin\ton{2 E + \omega}}
 } + \mathcal{O}\ton{e^2}.
\end{align}

Inserting $E = E_0 + 2\uppi$ into the full expression of $\Delta V\ton{E}$ allows to obtain the exact shift of the radial velocity per orbit, which is
\begin{align}
\left.\Delta V\right|_{2\uppi} \nonumber & = -\rp{\uppi\,\mathcal{K}\,a^{5/2}\,\sin I}{2\,\sqrt{\mu}\,\ton{1 - e\cos E_0}^3}\,\ton{
\sqrt{1 - e^2}\,\qua{-8 - 9\,e^2 + \right.\right.\\ \nonumber \\
\nonumber &\left.\left. + 6\,e\,\ton{-4\,\cos E_0  + e\,\cos 2 E_0}}\,\cos \omega\,\sin E_0  + \grf{-8\,\cos E_0  - 6\,e^4\,\cos^3 E_0  + \right.\right.\\ \nonumber \\
&\left.\left. + e\,\qua{2 + 18\,e^2 + 3\,\ton{-4 + e^2}\,\cos 2 E_0  + 3\,e\,\cos 3 E_0}}\,\sin \omega
}.
\end{align}

In the case of the Cosmological Constant $\Lambda$, it is
\eqi
\mathcal{K} = \rp{\Lambda\,c^2}{3}= 3.5\times 10^{-36}\,\mathrm{s^{-2}},
\eqf
where $c$ is the speed of light in vacuum.
In view of the small value of the Cosmological Constant, of the order of\footnote{$\Lambda$ can
be expressed in terms of the measurable parameters $H_0$ and $\Omega_\Lambda$, where $H_0$ is the Hubble parameter and $\Omega_\Lambda$ is the energy density of the Cosmological Constant normalized to the critical density. Their determinations from the measurements of the Cosmic Microwave Background (CMB) power spectra by the satellite Planck can be retrieved in \citet{2016A&A...594A..13P}.} \citep{2016A&A...594A..13P} $\Lambda\simeq 10^{-52}\,\mathrm{m}^{-2}$ and of the functional form of \rfr{Acc}, only very wide binaries for which RV's measurements exist \citep{1990AJ....100.1968C,2017A&A...598L...7K} could be, in principle, adopted to tentatively  constrain $\Lambda$. Since the available observational records do not cover an entire orbital period for such systems, the instantaneous expression of \rfr{DV1} has to be used to track the orbital arcs for which data exist.
By looking at the orbit of Proxima  about $\alpha$ Centauri AB, for which accurate RV's measurements exist \citep{2017A&A...598L...7K}, it is possible to infer an order of magnitude of the signal of \rfr{DV1} as little as
\eqi
\left|\Delta V_\Lambda\right|\lesssim 4\times 10^{-7}\,\mathrm{m\,s}^{-1}.
\eqf
The current accuracy in measuring the Proxima's RV is of the order of \citep{2017A&A...598L...7K}
\eqi
\sigma_{V}\simeq 30\,\mathrm{m\,s}^{-1};
\eqf
improvements of the order of a factor of two\footnote{P. Kervella, private communication, 2022.} may be obtained with new instruments such as ESPRESSO \citep{2021A&A...645A..96P} on the Very Large Telescope (VLT).
\section{ The case of a $r^{-3}$ perturbing acceleration}\lb{sec:4}
A perturbing acceleration in the form
\eqi
{\bds A}_{\mathcal{H}} = \rp{\mathcal{H}}{r^3}\, {\mathbf{u}}_R\lb{Acc2}
\eqf
arises in different models of  gravity. We remember that in General Relativity it corresponds to the contribute of the gravitational field due to a charged non-rotating spherically symmetric source in the Reissner-Nordstr\"om solution \citep{wald2010general}.  A similar perturbing acceleration is present in $f(T)$ gravity, for spherically symmetric solutions \citep{Iorio:2012cm,Ruggiero:2015oka}, in Einstein-Gauss-Bonnet gravity\citep{Maeda:2006hj,Bhattacharya:2016naa} and when one considers the quantum corrections to  the Schwarzschild solution\citep{Ali:2015tva,Jusufi:2016sym}, just to mention some examples.

If the extra-acceleration is given by \rfr{Acc2}, adopting the true anomaly $f$ is computationally more efficient. The following useful Keplerian expressions are used in the calculation
\begin{align}
r \lb{rf}&= \rp{p}{1+e\,\cos f}, \\ \nonumber \\
\dert{t}{f} \lb{dtdf}& = \rp{r^2}{\sqrt{\mu\,p}},
\end{align}
where $p\doteq a\,\ton{1-e^2}$ is the orbit's semilatus rectum.

From
\eqi
\Delta\kappa\ton{f} = \int_{f_0}^f\dert{\kappa}{t}\dert{t}{f^{'}}\mathrm{d}f^{'},\,\kappa=a,\,e,\,I,\,\Omega,\,\omega,\,\eta,\lb{dkdf}
\eqf
calculated with \rfrs{rf}{dtdf} and the Gauss equations for $\mathrm{d}\kappa/\mathrm{d}t,\,\kappa=a,\,e,\,I,\,\Omega,\,\omega,\,\eta$, one gets
\begin{align}
\Delta a\ton{f} \lb{Daf}& = \rp{\mathcal{H}\,e\,\ton{\cos f-\cos f_0}\,\qua{2 + e\,\ton{\cos f +\cos f_0}}}{\mu\,\ton{1-e^2}^2}, \\ \nonumber \\
\Delta e\ton{f} \lb{Def}& = \rp{\mathcal{H}\,\ton{\cos f-\cos f_0}\,\qua{2 + e\,\ton{\cos f +\cos f_0}}}{2\,\mu\,a\,\ton{1-e^2}}, \\ \nonumber \\
\Delta I\ton{f} \lb{DIf}& = 0, \\ \nonumber \\
\Delta\Omega\ton{f} \lb{DOf}& = 0, \\ \nonumber \\
\Delta\omega\ton{f} \lb{Dof}& = -\rp{\mathcal{H}\,\qua{3\,e\,\ton{-f + f_0} - \ton{2 + e\,\cos f}\,\sin f + \ton{2 + e\,\cos f_0}\,\sin f_0}}{2\,\mu\,a\,e\,\ton{1-e^2}}, \\ \nonumber \\
\Delta\eta\ton{f} \lb{Detaf}& = \rp{\mathcal{H}\,\qua{3\,e\,\ton{f - f_0} - \ton{2 + e\,\cos f}\,\sin f  + \ton{2 + e\,\cos f_0}\,\sin f_0 }}{2\,\mu\,a\,e\,\sqrt{1-e^2}}.
\end{align}
Furthermore, it is
\begin{align}
\int_{t_0}^t\Delta\nk\ton{t^{'}}\mathrm{d}t^{'} & \lb{Intdnf}\nonumber =\rp{3\,\mathcal{H}}{2\,\mu\,a\,\ton{1-e^2}^2\,\ton{1 + e\,\cos f}}\,\ton{
-\ton{1 - e^2}^{3/2}\,\ton{f - f_0}\,\ton{1 + e\cos f} -\right. \\ \nonumber \\
\nonumber &\left. - 2\,\arctan\ton{\rp{\ton{-1 + e}\,\tan\ton{\rp{f}{2}}}{\sqrt{1 - e^2}}}\,\ton{1 + e\cos f}\,\ton{1 + e\cos f_0}^2 + \right.\\ \nonumber \\
\nonumber &\left. + \ton{1 +
    e\cos f_0}\,\grf{2\,\arctan\ton{\rp{\ton{-1 + e}\,\tan\ton{\rp{f_0}{2}}}{\sqrt{1 - e^2}}}\,\ton{1 + e\cos f}\,\ton{1 + e\cos f_0} -\right.\right.\\ \nonumber \\
  &\left.\left. -  e\,\sqrt{1 - e^2}\,\qua{\sin f  + e\,\sin\ton{f - f_0} -\sin f_0}}
}.
\end{align}
Thus, inserting \rfrs{Detaf}{Intdnf} in \rfr{DMgeneral} yields
\begin{align}
\Delta\mathcal{M}\ton{f} \nonumber \lb{DMf}& = \rp{\mathcal{H}}{2\,\mu\,a\,\ton{1-e^2}^2}\,\ton{
-6\,\ton{1 + e\,\cos f_0}^2\,\arctan\ton{\rp{\ton{-1+e}\,\tan\ton{\rp{f}{2}}}{\sqrt{1-e^2}}} + \right.\\ \nonumber \\
\nonumber &\left. + 6\,\ton{1 + e\,\cos f_0}^2\,\arctan\ton{\rp{\ton{-1+e}\,\tan\ton{\rp{f_0}{2}}}{\sqrt{1-e^2}}} + \right.\\ \nonumber \\
\nonumber &\left. + \rp{1}{4\,e\,\ton{-1 + e^2 - \sqrt{1 - e^2}}\,\ton{1 + e\,\cos f}}\,\ton{-1 + e^2}\,\ton{1 + \sqrt{1 - e^2}}\,\grf{
-\qua{8 + 5\,e^2\,\ton{1 + e^2} + \right.\right.\right.\\ \nonumber \\
\nonumber &\left.\left.\left. + 6\,e^3\,\ton{4\,\cos f_0  + e\,\cos 2 f_0 }}\,\sin f  +
 e\,\ton{-1 + e^2}\,\ton{6\,\sin 2 f  + e\,\sin 3 f } + \right.\right.\\ \nonumber \\
&\left.\left. +  4\,\ton{2 + e^2}\,\ton{1 + e\,\cos f }\,\sin f_0  +
 2 e\,\ton{1 + 2 e^2}\,\ton{1 + e\,\cos f }\,\sin 2 f_0
}
}.
\end{align}
The full expression for the instantaneous shift of the RV due to \rfr{Acc2} can be obtained by inserting \rfrs{Daf}{Dof} and \rfr{DMf} in \rfrs{DvR}{DvN}, and taking the $z$ component ${\Delta\mathbf{v}}_z$ of the resulting velocity change $\Delta\mathbf{v}$; it is too cumbersome to be explicitly displayed. An expansion to the first power of $e$ of it is shown below:
\begin{equation}
\Delta V\ton{f} \nonumber \lb{DV2} = \rp{\mathcal{H}\,\sin I}{4\,\sqrt{\mu}\,a^{3/2}}\,\mathcal{V}\ton{f}\textcolor{black}{,}
\end{equation}
where
\begin{align}
\mathcal{V}\ton{f} \lb{DV2add}\nonumber & = \grf{
4\,\qua{\cos u - \cos\ton{2 f - f_0 + \omega}} + 8\,\ton{-f + f_0 }\,\sin u + \right.\\ \nonumber \\
\nonumber &\left.  +  e\,\qua{
\cos\ton{f - f_0 - \omega} + 9\,\cos \omega - 8\,\cos\ton{2 f + \omega} - \cos\ton{2 f - 2 f_0 + \omega} - \right.\right.\\ \nonumber \\
\nonumber & \left.\left. - 9\,\cos\ton{u - f_0} - \cos\ton{3 f - f_0 + \omega} + 9\,\cos\ton{u + f_0} + 2\,\ton{-f + f_0}\,\sin\omega - \right.\right.\\ \nonumber \\
&\left.\left. - 12\,\ton{f - f_0}\,\ton{\cos f +\,\cos f_0}\,\sin u
}
} + \mathcal{O}\ton{e^2}.
\end{align}
The  exact shift of the radial velocity per orbit can be obtained by inserting $f =  f_0 + 2\uppi$ into the full expression of $\Delta V\ton{f}$; it reads
\eqi
\left.\Delta V\right|_{2\uppi} = -\rp{\uppi\,\mathcal{H}\,\sin I\,\ton{e\,\sin\omega + \sin u_0}}{a^{3/2}\,\ton{1-e^2}^{3/2}\,\sqrt{\mu}}.\lb{DV2full}
\eqf

By assuming, e.g., $\mu_\star = 0.5\,\mu_\odot,\,\mu_\mathrm{p}=50\,\mu_\mathrm{Jup},\,a=0.002\,\mathrm{au},\,I=50^\circ$ along with an experimental uncertainty in measuring the RV as little as $\sigma_V\simeq 0.1\, \mathrm{m\,s}^{-1}$, \rfr{DV2full}
yields
\eqi
\left|\mathcal{H}\right|\lesssim 2\times 10^{22}\,\mathrm{m}^4\,\mathrm{s}^{-2}.\lb{bound2}
\eqf
The bound of \rfr{bound2} should be viewed just as preliminarily indicative of the possibility offered by the proposed approach, not as an actually obtainable constraint. To this aim, a detailed error budget including also the impact of several systematic errors like other competing dynamical effects should be assessed, along with the choice of the initial value of $u_0$ in order to maximize the signal-to-noise ratio.
\section{ The case of a $r^{-4}$ perturbing acceleration}\lb{sec:5}
In different models of gravity it is possible to obtain a perturbing acceleration in the form
\eqi
{\bds A}_{\mathcal{H}} = \rp{\mathcal{Q}}{r^4}\, {\mathbf{u}}_R.\lb{Acc3}
\eqf
Just to refer to some examples, we mention the  modification of the Schwarzschild solution obtained using the renormalization group approach \citep{Bonanno:2000ep}; the Sotiriou-Zhou solution  which is obtained starting from the coupling of a scalar field $\phi$ with the Gauss-Bonnet invariant, even though it cannot be used to describe the spacetime around a star but, rather, around a black hole \citep{Sotiriou:2014pfa,Bhattacharya:2016naa,Kanti:2011jz,Kanti:2011yv}; such a perturbing acceleration  arises also in the framework of string theory, in presence of the Kalb-Ramond field \citep{Chakraborty:2016lxo}. Eventually, it is important to remember that a perturbing acceleration in the form (\ref{Acc3}) arises also in other models of gravity which are effective at the scale of elementary particles, so they cannot be considered for our purposes (see e.g. \citet{Iorio:2018fam} and references therein).

Also in the case of \rfr{Acc3}, using the true anomaly $f$ is computationally more efficient.
Inserting \rfr{Acc3} in \rfr{dkdf} and using \rfrs{rf}{dtdf} yields
\begin{align}
\Delta a\ton{f} \lb{Da3}\nonumber & = -\rp{2\,\mathcal{Q}\,e}{\mu\,a\,\ton{1-e^2}^3}\,\grf{-\rp{1}{3}\,\cos f\,\qua{3 + e\,\cos f\,\ton{3 + e\,\cos f}} +\,\cos f_0 + \right.\\ \nonumber \\
 &\left. + e\,\cos^2 f_0 + \rp{1}{3}\,e^2\,\cos^3 f_0}, \\ \nonumber \\
\Delta e\ton{f} \lb{De3}\nonumber & = \rp{\mathcal{Q}}{3\,\mu\,a^2\,\ton{1-e^2}^2}\,\grf{
3\,\cos f  + 3\,e\,\cos^2 f  + e^2\,\cos^3 f  - \right.\\ \nonumber \\
&\left. - \cos f_0\,\qua{3 + e\,\cos f_0\,\ton{3 + e\,\cos f_0 }}
}, \\ \nonumber \\
\Delta I\ton{f} \lb{DI3}& = 0, \\ \nonumber \\
\Delta \Omega\ton{f} \lb{DO3}& = 0, \\ \nonumber \\
\Delta \omega\ton{f} \lb{Do3}\nonumber & = \rp{\mathcal{Q}}{12\,\mu\,a^2\,e\,\ton{1-e^2}^2}\,\grf{
3\,\ton{4 + 3 e^2}\,\sin f  + e\,\qua{6\,\ton{2 f - 2 f_0 +\,\sin 2 f } + e\,\sin 3 f } - \right.\\ \nonumber \\
&\left. - 2\,\ton{6 + 5 e^2 + 6 e\,\cos f_0  + e^2\,\cos 2 f_0 }\,\sin f_0
}, \\ \nonumber \\
\Delta \eta\ton{f} \lb{Deta3}\nonumber & = -\rp{\mathcal{Q}}{12\,\mu\,a^2\,\ton{1-e^2}^{3/2}}\,\ton{
3\,\ton{4 - 5 e^2}\,\sin f  - 12\,\qua{e\,\ton{f - f_0} + \sin f_0 } + \right. \\ \nonumber \\
&\left. + e\,\grf{6\,\sin 2 f  + e\,\sin 3 f  - 2\,\qua{6\,\cos f_0  + e\,\ton{-7 + \cos 2 f_0 }}\,\sin f_0 }\textcolor{black}{.}
}
\end{align}
\Rfr{konz} allows to obtain
\begin{align}
\int_{t_0}^t\Delta\nk\ton{t^{'}}\mathrm{d}t^{'} & \lb{Intdnf3}\nonumber = -\rp{\mathcal{Q}}{4\,\mu\,a^2\,\ton{1-e^2}^3\,\ton{1 + e\,\cos f}}\,\ton{
4\,\ton{1 - e^2}^{3/2}\,\ton{f - f_0}\,\ton{1 + e\,\cos f} + \right.\\ \nonumber \\
\nonumber &\left. + \ton{8\arctan\ton{\rp{\ton{- 1 + e}\tan\ton{\rp{f}{2}}}{\sqrt{1-e^2}}} - \right.\right.\\ \nonumber \\
\nonumber &\left.\left. - 8\arctan\ton{\rp{\ton{- 1 + e}\tan\ton{\rp{f_0}{2}}}{\sqrt{1-e^2}}}}\,\ton{1 +
e\,\cos f}\,\ton{1 + e\,\cos f_0}^3 +\right.\\ \nonumber \\
\nonumber &\left. + e\,\sqrt{1 - e^2}\,\ton{2 \ton{4 + e^2 +
      e\,\ton{-2 \ton{-1 + e^2}\,\cos f  + 6\,\cos f_0  + \right.\right.\right.\right.\\ \nonumber \\
\nonumber &\left.\left.\left.\left. +      2\,e^2\,\cos^3  f_0  +
         3\,e\,\cos 2 f_0 }}\,\sin f  -
   2 \ton{1 + e\,\cos f } \ton{4 - e^2 + \right.\right.\right.\\ \nonumber \\
&\left.\left.\left. +  4\,e\,\cos f_0  + e^2\,\cos 2 f_0 }\,\sin f_0 }
}.
\end{align}
The shift of the mean anomaly, computed with \rfr{Deta3} and \rfr{Intdnf3}, turns out to be
\begin{align}
\Delta\mathcal{M}\ton{f} \lb{DM3} \nonumber & = \rp{\mathcal{Q}}{12\,\mu\,a^2}\,\ton{
-\rp{1}{e\,\ton{1-e^2}^{3/2}}\,\ton{3\,\ton{4 - 5 e^2}\,\sin f - 12\,\ton{e\,\ton{f - f_0} +\,\sin f_0} + \right.\right.\\ \nonumber \\
\nonumber &\left.\left. + e\,\ton{6\,\sin 2 f + e\,\sin 3 f - 2\,\ton{6\,\cos f_0 + e\,\ton{-7 + \cos 2 f_0}}\,\sin f_0} } -\right.\\ \nonumber \\
\nonumber &\left. -\rp{3}{\ton{1-e^2}^3\,\ton{1+e\,\cos f}}\,\ton{
4\,\ton{1 - e^2}^{3/2}\,\ton{f - f_0}\,\ton{1 + e\,\cos f} +\right.\right.\\ \nonumber \\
\nonumber &\left.\left. +  \ton{1 + e\,\cos f}\,\ton{1 + e\,\cos f_0}^3\,\ton{8\,\arctan\ton{\rp{\ton{-1+e}\,\tan\ton{\rp{f}{2}}}{\sqrt{1-e^2}}}-\right.\right.\right.\\ \nonumber \\
\nonumber &\left.\left.\left. - 8\,\arctan\ton{\rp{\ton{-1+e}\,\tan\ton{\rp{f_0}{2}}}{\sqrt{1-e^2}}}
} + e\,\sqrt{1-e^2}\,\ton{
2\,\ton{4 + e^2 + e\,\ton{-2\,\ton{-1 + e^2}\,\cos f  + \right.\right.\right.\right.\right.\\ \nonumber \\
\nonumber &\left.\left.\left.\left.\left. + 6\,\cos f_0  + 2 e^2\,\cos f_0 ^3 + 3 e\,\cos 2 f_0 }}\,\sin f  - \right.\right.\right.\\ \nonumber \\
&\left.\left.\left. -  2\,\ton{1 + e\,\cos f }\,\ton{4 - e^2 + 4 e\,\cos f_0  + e^2\,\cos 2 f_0 }\,\sin f_0
}
}
}.
\end{align}

Thus, \rfrs{DvR}{DvN}, calculated with \rfrs{Da3}{Deta3} and \rfr{DM3}, provide the change in the radial velocity due to \rfr{Acc3}; it is
\begin{equation}
\Delta V\ton{f} \nonumber \lb{DV3} = \rp{\mathcal{Q}\,\sin I}{a^{5/2}\,\sqrt{\mu}} \mathcal{V}\ton{f}\textcolor{black}{,}
\end{equation}
where
\begin{align}
 \mathcal{V}\ton{f}& \lb{DV3add} = \ton{
\cos u - \cos\ton{2 f - f_0 + \omega} + \ton{-f + f_0 - \right.\right.\\ \nonumber \\
\nonumber & -\left.\left. 2\,\arctan\ton{\tan\ton{\rp{f}{2}}} + 2\,\arctan\ton{\tan\ton{\rp{f_0}{2}}}}\,\sin u + \right.\\ \nonumber \\
\nonumber & +\left. \rp{e}{4}\,\ton{
\cos\ton{f - f_0 - \omega}  + 6\,\cos \omega  - 4\,\cos\ton{2 f + \omega}  - 2\,\cos\ton{2 f - 2 f_0 + \omega}  - \right.\right.\\ \nonumber \\
\nonumber &\left.\left. - 5\,\cos\ton{f - f_0 + \omega}  -\cos\ton{3 f - f_0 + \omega}  + 5\,\cos\ton{f + f_0 + \omega}  + \right.\right.\\ \nonumber \\
&\left.\left. + 4\,\ton{-f + f_0}\,\sin \omega - 8\,\ton{f - f_0}\,\ton{2\,\cos f + 3\,\cos f_0}\,\sin u
}
}+\mathcal{O}\ton{e^2}.
\end{align}
%

By putting $f = f_0 + 2\uppi$ in the full expression of $\Delta V\ton{f}$, one gets the exact shift of the radial velocity per orbit:
\eqi
\left.\Delta V\right|_{2\uppi} = \rp{2\uppi\,\mathcal{Q}\,\sin I\,\ton{e\,\sin\omega + \sin u_0}}{a^{5/2}\,\ton{1-e^2}^{5/2}\,\sqrt{\mu}}.\lb{DV3full}
\eqf

By adopting  the same values for the exoplanet's RV uncertainty and physical and orbital parameters as in Section\,\ref{sec:4},
\rfr{DV3full} returns
\eqi
\left|\mathcal{Q}\right|\lesssim 5\times 10^{30}\,\mathrm{m^5\,s^{-2}}.\lb{bound3}
\eqf
As \rfr{bound2}, also the bound of \rfr{bound3} should be viewed as preliminarily and just indicative of the potential of the proposed strategy.
\section{Summary and conclusions}\lb{sec:6}
We analytically calculated the shifts $\Delta V$ of the RV $V$ of a gravitationally bound binary system induced by some long-range modified models of gravity which could be tested or constrained with existing RV data of known binaries such as, e.g., exoplanets and binary stars.
In particular, exoplanets close to their parents stars for which long RV records covering several orbital revolutions are available can constrain power-law models yielding extra-accelerations proportional to $r^{-N-1},\,N=2,\,3$.; in this case, the expressions for the RV changes over one orbital period are to be used. It is useful to recall  that in a previous paper \citep{Iorio:2018fam} we constrained the extra acceleration parameters using existing long data records of the LAGEOS satellites, tracked on an almost continuous basis with the Satellite Laser Ranging (SLR) technique.  In that case, we obtained the following constraints $\displaystyle \left|\mathcal{H}_\mathrm{SLR}\right|\lesssim 4.2\times 10^{6}\,\mathrm{m}^4\,\mathrm{s}^{-2}$ in the $N=2$ case, and $\displaystyle \left|\mathcal{Q}_\mathrm{SLR}\right|\lesssim 1.2 \times 10^{13}\,\mathrm{m^5\,s^{-2}}$ in the $N=3$ case.
\textcolor{black}{Other preliminary constraints on $\mathcal{Q}$ came from some astronomical and astrophysical systems; see Table 2 of \citet{IorioAnP2012}. The spacecraft GRACE provided $\displaystyle \left|\mathcal{Q}_\mathrm{GRACE}\right|\lesssim 1\times 10^{16}\,\mathrm{m^5\,s^{-2}}$, while the pericentre precessions of the Solar System's planets and of the S2 star around Sagittarius A$^\ast$ yielded $\displaystyle \left|\mathcal{Q}_\mathrm{SS}\right|\lesssim 1.5\times 10^{29}\,\mathrm{m^5\,s^{-2}}$ and $\displaystyle \left|\mathcal{Q}_\mathrm{Sgr\,A^{\ast}}\right|\lesssim 5\times 10^{50}\,\mathrm{m^5\,s^{-2}}$, respectively.}
Accordingly, the constraints that can be obtained using the method described in this paper are much looser, and the motivation is evident: the perturbing accelerations in the form (\ref{Acc2}) and (\ref{Acc3}) are rapidly decreasing, so that it is more effective to test them on smaller scales, such as those of satellites orbiting the Earth, rather than at planetary system scale.

On the other hand, wide binaries with orbital periods of the order of hundreds of thousands of years like Proxima and $\alpha$ Centauri AB may, in principle, be used to put constraints on Hooke-type anomalous accelerations proportional to $r$; the CC belongs to such a family of models. In such scenarios, RV data necessarily cover just relatively short orbital arcs; thus, the instantaneous expressions for $\Delta V$ must be adopted. In the particular case of the CC, it induces a RV shift of the Proxima/$\alpha$ Centauri AB system as little as $\left|\Delta V\right|\lesssim 10^{-7}\,\mathrm{m\,s}^{-1}$, while the current accuracy in measuring its RV is of the order of $\sigma_V\simeq 30\,\mathrm{m\,s}^{-1}$.

Nonetheless, we believe that our approach could be useful to test the prediction of modified models of gravity outside the Solar System. In addition, we point out that the calculational scheme set up in this work is completely general, and can be straightforwardly extended to other modified models of gravity for which explicit expressions for the resulting extra-accelerations are available.
%
\bibliography{exopbib,PXbib}{}

\end{document}